\documentclass[fleqn,twoside]{article}

\usepackage{espcrc2}
\usepackage{slashed}
\usepackage{slashbox}
\usepackage{multirow}
\usepackage{graphicx}
\usepackage[dvips]{color}
\usepackage{epsfig,psfig}
\usepackage{colortbl}
\renewcommand{\d}{\textrm{d}}
\newcommand{\der}[2]{\frac{{\d}#1}{{\d}#2}}   

\newcommand{\sigva}[0]{\left\langle \sigma_{{A}}v\right\rangle}

\def\alt{\raise0.3ex\hbox{$\;<$\kern-0.75em\raise-1.1ex\hbox{$\sim\;$}}}
\def\agt{\raise0.3ex\hbox{$\;>$\kern-0.75em\raise-1.1ex\hbox{$\sim\;$}}}

\definecolor{Black}{named}{Black}
\definecolor{Red}{named}{Red}

\newcommand{\bw}{\begin{widetext}}
\newcommand{\ew}{\end{widetext}}
\def\d{{\rm d}}

\begin{document}
\title{Radio Signal by Galactic Dark Matter}
\author{E. Borriello\address[UNINA]{Universit\`a di Napoli ``Federico II'', Dipartimento di
Scienze Fisiche and INFN - Sezione di Napoli, Complesso
Universitario di Monte S.Angelo, Via Cithia, 80126, Napoli,
Italy}, A. Cuoco\address[DK]{Department of Physics and Astronomy,
                University of Aarhus, Ny Munkegade, Bygn. 1520 8000 Aarhus Denmark}, G. Miele\addressmark[UNINA]
\address[IFIC]{Instituto de F\'{\i}sica Corpuscular (CSIC-Universitat de
Val\`encia),
        Ed.\ Institutos de Investigaci\'on, Apartado de Correos 22085, E-46071 Val\`encia,
        Spain.}\thanks{Presented by G. Miele}
                }

\begin{abstract}
An interesting strategy for indirect detection of Dark Matter
comes through the amounts of electrons and positrons usually
emitted by DM pair annihilation. The $e^+e^-$ gyrating in the
galactic magnetic field then produce secondary synchrotron
radiation. The radio emission from the galactic halo as well as
from its expected substructures if compared with the measured
diffuse radio background can provide constraints on the physics of
WIMPs. In particular one gets the bound of $\sigva=
10^{-24}$\,cm$^3$s$^{-1}$ for a DM mass $m_{\chi}=100$\,GeV even
though sensibly depending on the astrophysical uncertainties.

\end{abstract}
\maketitle

\section{Introduction}
Among the indirect DM detection channels, the radio emission due
to secondary electrons or positrons can represent a chance to look
for DM annihilation. During the process of thermalization in the
galactic medium the high energy $e^+$ and $e^-$ release secondary
low energy radiation, in particular in the radio and X-ray band,
which in principle could be detected. Furthermore, while the
astrophysical uncertainties affecting this signal are similar to
the case of direct $e^+$, $e^-$ detection, the sensitivities are
quite different, and, in particular, the radio band allows for a
the discrimination of tiny signals even in a background many order
of magnitudes more intense.

Indirect detection of DM annihilation through secondary photons
has received recently an increasing attention, exploring the
expected signature in X-rays
\cite{Bergstrom:2006ny,Regis:2008ij,Jeltema:2008ax}, at radio
wavelengths
\cite{Blasi:2002ct,Aloisio:2004hy,Tasitsiomi:2003vw,Zhang:2008rs}
, or both
\cite{Colafrancesco:2005ji,Colafrancesco:2006he,Baltz:2004bb}. In
this paper we will focus our analysis on the radio signal expected
from the Milky Way (MW) halo and its substructures. The results
and the details of the following approach can be found in Ref.
\cite{Borriello:2008gy}.

\section{Dark matter distribution}\label{DMD}

Our knowledge of the DM spatial distribution on galactic and
subgalactic scales has greatly improved thanks to recent high
resolution zoomed N-body simulations
\cite{Diemand:2005vz,Diemand:2006ik,Kuhlen:2008aw,Springel:2008cc}.
These simulations indicate that for the  radial profile of the
galactic halo the usual Navarro-Frank-White (NFW) distribution
\cite{Navarro:1996gj}
\begin{equation}
    \label{NFW}
    \rho(r)=\frac{\rho_{{h}}}{\frac{r}{r_{{h}}}\left(1+\frac{r}{r_{{h}}}\right)^2}
    \,\,\, ,
\end{equation}
still works as a good approximation over all the resolved scales.
The NFW profile is in fair agreement with the observed Milky Way
rotation curve \cite{Klypin:2001xu}, although, depending on the
employed model, it is possible to find an agreement for many
different DM profiles. We emphasize, however, that the various
profiles differ mainly in the halo center (for $r\alt 1$ kpc)
where the uncertainties, both in numerical simulations and from
astrophysical observations are maximal. Thus, our analysis which
explicitly excludes the galactic center, does not crucially depend
on the choice of the profile.

The parameters describing the halo are then determined imposing
the DM density to be equal to $\rho_S=0.365$ GeV $c^{-2}$
cm$^{-3}$ near the Solar System, at a galactocentric distance of
$R_{{S}}=8.5$ kpc.

Simulations, however, predict a DM distribution sum of a smooth
halo component, and of an additional clumpy one with total masses
roughly of the same order of magnitude. For the subhalo population
we will assume a mass distribution $\propto m_{{cl}}^{-2}$ and
that they are spatially distributed following the NFW profile of
the main halo. The mass spectrum number density of subhaloes, in
galactocentric coordinates $\vec{r}$, is thus given by
\begin{equation}
\frac{{\d} n_{{cl}}}{{\d} m_{{cl}}}(m_{{cl}},\vec{r})= A \left(
\frac{m_{{cl}}}{M_{{cl}}} \right)^{-2} \left( \frac{r}{r_{{h}}}
\right)^{-1}\left(1+\frac{r}{r_{{h}}}\right)^{-2},
\end{equation}
where $A$ is a dimensional normalization constant.

By using the constraints described in \cite{Borriello:2008gy} one
can fix the values of free parameters $r_{{h}}$, $\rho_{{h}}$ and
$A$, hence obtaining $r_{{h}}=14.0$ kpc, $\rho_{{h}}=0.572$\,GeV
$c^{-2}$ cm$^{-3}$ and $A=1.16 \times 10^{-19}$ kpc$^{-3}$
M$_{\odot}^{-1}$.

A further piece of information is required to derive the annihilation
signal from the clumps, namely how the DM is distributed inside the
clumps themselves. We will assume that each clump follows a NFW
profile whose parameters are fixed assuming the clumps concentration
according to \cite{Bullock:1999he}.

\section{Radio Data}
In the following we will derive  constraints on the DM emission
comparing the expected diffuse emission from the smooth halo and
the unresolved population of clumps with all sky observation in
the radio band. In the frequency range 100 MHz-100 GHz where the
DM synchrotron signal is expected, various astrophysical processes
contribute to the observed diffuse emission. Competing synchrotron
emission is given by Cosmic Ray electrons accelerated in
supernovae shocks dominating the radio sky up to $\sim$ 10 GHz. At
higher frequencies the Cosmic Microwave Background (CMB) and its
anisotropies represent the main signal. However, thanks to the
very sensitive multi-frequency survey by the WMAP satellite, this
signal (which represents thus a background for DM searches) can be
modeled in a detailed way and can thus be removed from the
observed radio galactic emission \cite{Tegmark:2003ve}. Other
processes contributing in the 10-100 GHz range are given by
thermal bremsstrahlung (free-free emission) of electrons on the
galactic ionized gas, and emission by small grains of vibrating or
spinning dust.

In the following our approach will be  to compare the DM signal
with the observed radio emission where only the CMB is modeled and
removed. For this purpose we use the code described in
\cite{deOliveiraCosta:2008pb} where most of the radio survey
observations in the range 10 MHz-100 GHz are collected and a
scheme to derive interpolated, CMB cleaned sky maps at any
frequency in this range is described.

\section{DM Synchrotron Signal}\label{synchsignal}

In a standard scenario where WIMPs experience a non exotic thermal
history, a typical mass range for these particles is $ 50 \,
\textrm{GeV} \alt m_\chi \alt 1 \, \textrm{TeV}$, while a simple
estimate for their (thermally averaged) annihilation cross section
yields $\sigva = 3 \times 10^{-27} \textrm{cm}^3\textrm{s}^{-1}/
\, \Omega_{{cdm}} h^2$, giving $\sigva \approx 3 \times 10^{-26}
\textrm{cm}^3\textrm{s}^{-1}$ for $\Omega_{{cdm}} h^2 \approx0.1$
as resulting from the latest WMAP measurements. However, this
naive relation can fail badly if, for example, coannihilations
play a role in the WIMP thermalization process
\cite{Griest:1990kh}, and a much wider range of cross sections
should be considered viable. In this work we consider values of
$m_\chi$ from about 10 GeV to about 1 TeV, and $\sigva$ in the
range $10^{-26}\div10^{-21}\,\textrm{cm}^3\textrm{s}^{-1}$

The $e^+e^-$ annihilation spectrum, ${\d} N_e/ {\d}E_e(E_e)$, for a given
super-symmetric WIMP candidate can be calculated for example with
the DarkSUSY package \cite{Gondolo:2004sc}.  Once injected in the
galaxy, by neglecting diffusion (see Ref. \cite{Borriello:2008gy}
for details), the emitted $e^+e^-$ follow the steady state
distribution which reads
\begin{equation}\label{steady}
\der{n_e}{E_e}(E_{e},\vec{r}) = \frac{\tau(\vec{r})}{E_{e}}
\int_{E_{e}}^{m_\chi c^2} \!\!\!\!\! dE_{e}' \,\,
Q(E_{e}',r)\, ,
\end{equation}
where
\begin{equation} \label{inj1}
Q(E_e,r)=\frac{1}{2}\left(\frac{\rho(r)}{m_{\chi}}\right)^2 \sigva
\der{N_e}{E_e}(E_e)\
\end{equation}
is the constant rate at which DM annihilation injects new electrons in the galaxy
and $\tau=E_e/b(E_e,\vec{r})$ is the cooling time, resulting
from the sum of several energy loss processes that affect
electrons. In the following we will consider synchrotron emission
and Inverse Compton Scattering (ICS) off the background photons
(CMB and starlight) only, which are the faster processes and thus
the ones really driving the electrons equilibrium. Other
processes, like synchrotron self absorption, ICS off the
synchrotron photons, $e^+ e^-$ annihilation, Coulomb scattering
over the galactic gas and bremsstrahlung are generally slower.
They can become relevant for extremely intense magnetic field,
possibly present in the inner parsecs of the galaxy
\cite{Aloisio:2004hy}, and thus will be neglected in this
analysis.

The synchrotron spectrum of an electron gyrating in a magnetic
field has its prominent peak at the resonance frequency
\begin{equation}
    \label{peaknu}
\nu=3.7\left(\frac{B}{\mu
\textrm{G}}\right)\left(\frac{E_e}{\textrm{GeV}}\right)^2 \textrm{MHz}\ .
\end{equation}
This implies that, in
practice, a $\delta$--approximation around the peaks works
extremely well. Using this frequency peak approximation,
the synchrotron emissivity can be defined as
\begin{equation}
    \label{synemiss}
j_{\nu}(\nu,\vec{r})
                =\der{n_e}{E_e}(E_e(\nu),\vec{r})\,\der{E_e(\nu)}{\nu} b_{{syn}}(E_e(\nu),\vec{r}).
\end{equation}
This quantity is then integrated along the line of sight
 to get the final synchrotron flux across the sky:

\begin{equation}
\frac{\d^2 I_{\nu}}{{\d}l\,{\d}b}=\frac{\cos{b}}{4\pi}
\int_0^{\infty} j_{\nu} \, {\d}s \, ,
\end{equation}
where $(l,b)$ are coordinates on the sphere and $s$ the line of
sight coordinate.

The two contributes mainly differ due to the fact that, while the
halo emission is proportional to the square of the NFW profile,
the clumps one is simply proportional to it. (see Ref.
\cite{Borriello:2008gy}).

\begin{figure}[!t]
    \includegraphics[width=0.60\columnwidth, angle=90]{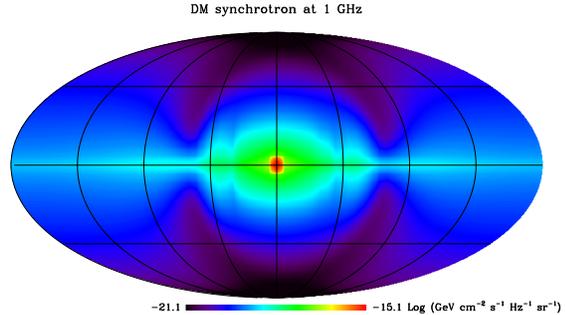}
    \caption{Sky map of the galactic radio signal generated by the DM smooth
    halo and  unresolved clumps
    at the frequency of 1\,GHz for $m_\chi=100$ GeV and
$\sigva=3\times10^{-26}$cm$^3$s$^{-1}$. The peculiar shape of the
signal, pinched approximately at $\pm 30^\circ$ and $\pm
60^\circ$, reflects basically the structure of the magnetic field
as seen in projection from the Solar System, where the observer is
located.}
  \label{fig:halocontours}
\end{figure}

\begin{figure}[!t]
 \centering
 \includegraphics[width=0.85\columnwidth]{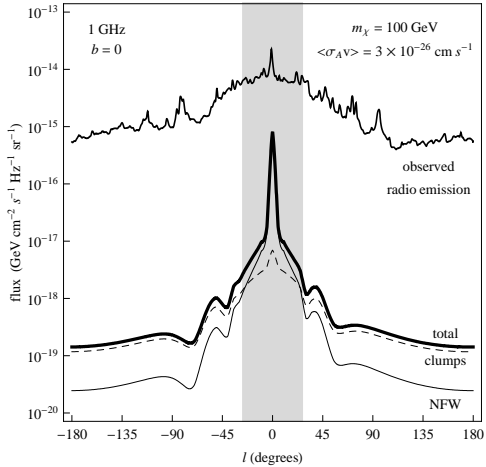}
 \caption{DM synchrotron profile for the halo and unresolved
 substructures and their sum at 1 GHz for $m_\chi=100$ GeV  and
 $\sigva=3\times10^{-26}$\,cm$^3$s$^{-1}$. The astrophysical observed
 emission at the same frequency is also shown. The gray band indicates
 the angular region within which the DM signal from the host halo dominates
 over the signal from substructures modeled as in section \ref{DMD}.}
 \label{fig:HVSCL}
\end{figure}

The halo component dominates in the central region of
the galaxy, inside a disk with a radius of about 30 degrees,
while the clumps component represents the main contribution in
the external region (see fig. \ref{fig:HVSCL}).

\section{DM Annihilation constraints}
The pattern and intensity of the DM radio map resulting from the
sum of the contributions from the smooth halo and unresolved
clumps is shown in Fig. \ref{fig:halocontours} for $m_\chi=100$
GeV and $\sigva=3\times10^{-26}$cm$^3$s$^{-1}$. Similar maps are
obtained at different frequencies and different $m_\chi$ and
$\sigva$ to obtain DM exclusion plots. For our analysis we use a
small mask covering a $15^\circ$$\times$$15^\circ$ region around
the galactic center where
 energy loss processes other than synchrotron and ICS start
possibly to be relevant. We include the galactic plane although
this region  has basically no influence for the constraints on the
DM signal.

\begin{figure*}[!t]
\begin{center}
\begin{tabular}{cc}
\includegraphics[width=.90\columnwidth]{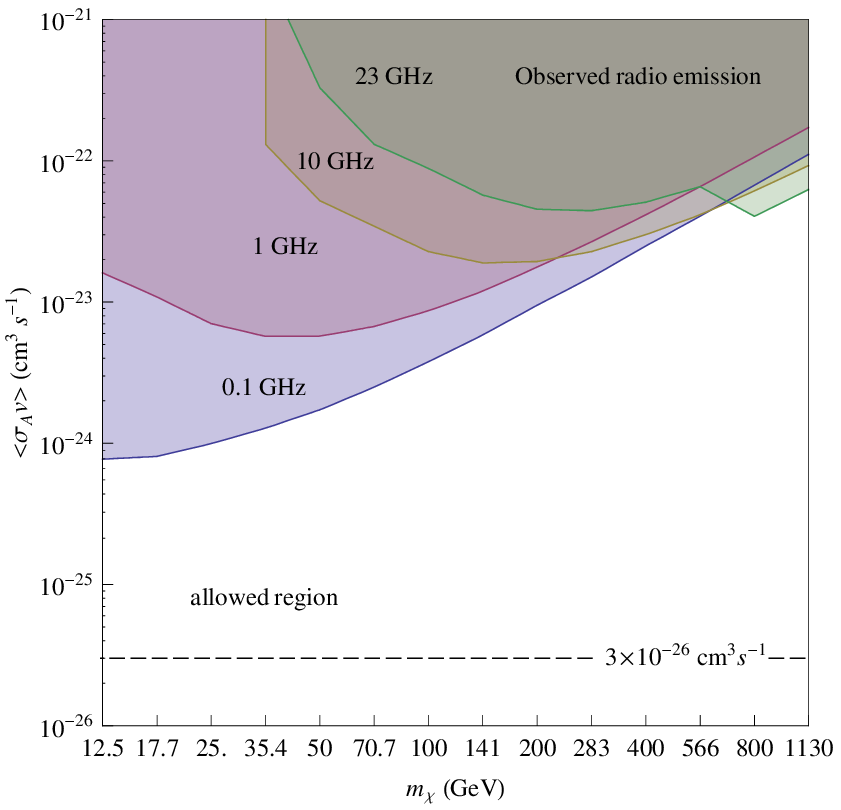} &
\hspace{1pc}
\includegraphics[width=.90\columnwidth]{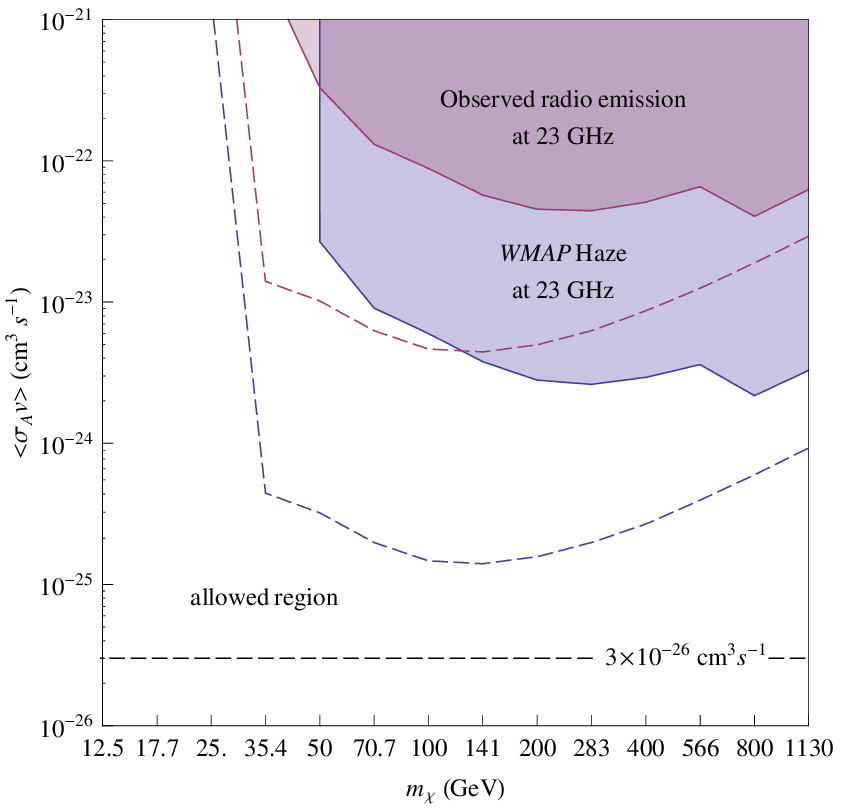}\\
\end{tabular}
\end{center}
\caption{(Left) Constraints in the  $m_\chi$-$\sigva$ plane for
various frequencies without assuming synchrotron foreground
removal. (Right) Constraints from the WMAP 23 GHz foreground map
and 23 GHz foreground--cleaned residual map (the WMAP Haze) for
the TT model of magnetic field (filled regions) and for a uniform
10 $\mu$G field (dashed lines). } \label{fig:DMconstraints}
\end{figure*}

In Fig.\ref{fig:DMconstraints} we show the radio constraints on
the DM annihilation signal in the $m_\chi$--$\sigva$ plane for
various frequencies and various choices of the foreground. Several
comments are in order. First, we can see that, as expected, the
use of the haze at 23 GHz gives about one order of magnitude
better constraints with respect to the synchrotron foregrounds at
the same frequency. However, using also the information at other
frequencies almost the same constraints can be achieved. This
information in particular is complementary giving better
constraints at lower DM masses. This is easily understood since a
smaller DM mass increases the annihilation signal ($\propto
m_\chi^{-2}$) at smaller energies, and thus smaller synchrotron
frequencies. In particular, the constraints improve of about one
order of magnitude at $m_\chi\sim100$ GeV from 23 GHz to 1 GHz
while only a modest improvement is achieved considering further
lower frequencies as 0.1 GHz. This saturation of the constraints
is due to the frequency dependence of the DM signal, that below 1
GHz becomes flatter than the astrophysical backgrounds so that the
fraction of contribution from DM is maximal at about 1 GHz.

For low masses the
constraints come more and more from lower frequencies. For example
for a WIMP of 30 GeV the data at 100 MHz are 2 orders of magnitude
more constraining than the data at 10 GHz. For a WIMP of 1 GeV,
from Eq.\ref{peaknu} with a magnetic field of
$\mathcal{O}$($\mu$G) only frequencies $\alt 10$ MHz would be
useful to place constraints on the DM signal. Although
observations at this frequency exist
\cite{deOliveiraCosta:2008pb}, in general the survey sky coverage
is quite incomplete and the data quality is non-optimal.
Observations in this very low frequency range should substantially
improve with the next generation radio arrays LOFAR and SKA.

\section{Summary and conclusions}
Using conservative assumptions for the DM distribution in our
galaxy we  derive the expected secondary radiation due to
synchrotron emission from high energy electrons produced in DM
annihilation. The signal from single bright clumps offers only
poor sensitivities because of diffusion effects which spread the
electrons over large areas diluting the radio signal. The diffuse
signal from the halo and the unresolved clumps is instead
relevant and can be compared to the radio astrophysical background
to derive constraints on the DM mass and annihilation cross
section.

Constraints in the radio band, in particular, are complementary to
similar (less stringent but less model dependent) constraints in
the X-ray/gamma band \cite{Mack:2008wu,Kachelriess:2007aj} and
from neutrinos \cite{Yuksel:2007ac}. Radio data, in particular,
are more sensitive in the GeV-TeV region while neutrinos provide
more stringent bounds for very high DM masses ($\agt 10$ TeV).
Gammas, instead, are more constraining for $m_\chi \alt 1$ GeV.
The combination of the various observations provides thus
interesting constraints over a wide range of masses  pushing the
allowed window significantly near the thermal relic possibility.

More into details, we obtain conservative constraints at the level
of $\sigva \sim 10^{-23}$\,cm$^3$s$^{-1}$ for a DM mass
$m_{\chi}=100$\,GeV from the WMAP Haze at 23 GHz. However,
depending on the astrophysical uncertainties, in particular on the
assumption on the galactic magnetic field model, constraints as
strong as $\sigva \sim 10^{-25}$\,cm$^3$s$^{-1}$ can be achieved.
Complementary to other works which employ the WMAP Haze at 23 GHz,
we also use  the information in a wide frequency band in the range
100 MHz-100 GHz. Adding this information the constraints become of
the order of $\sigva \sim 10^{-24}$\,cm$^3$s$^{-1}$ for a DM mass
$m_{\chi}=100$\,GeV. The multi-frequency approach thus gives
comparable constraints with respect to the WMAP Haze only, or
generally better for $m_\chi\alt100$ GeV where the best
sensitivity is achieved at $\sim$ GHz frequencies.

The derived constraints are quite conservative because no attempt
to model the astrophysical background is made differently from the
case of the WMAP Haze. Indeed, the Haze residual map itself should
be interpreted with some caution, given that the significance of
the feature is at the moment still debated and complementary
analyses from different groups (as the WMAP one) miss in finding a
clear evidence of the feature. Definitely the multi-frequency
approach will be necessary to test in a convincing way a possible
DM signal like the claim related to the WMAP Haze. Progresses are
expected with the forthcoming data at high frequencies from Planck
and at low frequencies from LOFAR and, in a more distant future,
from SKA. These surveys will help in disentangling the various
astrophysical contributions thus assessing the real significance
of the Haze feature. Further, the low frequency data in
particular, will help to improve our knowledge of the galactic
magnetic field. Progresses in these fields will provide a major
improvement for the interpretation of the DM-radio connection.

\vspace{1pc}
\section*{Acknowledgments}
G.M. acknowledges supports by the Spanish MICINN (grants
SAB2006-0171 and FPA2005-01269) and by INFN--I.S.Fa51 and PRIN
2006 ``Fisica Astroparticellare: Neutrini ed Universo Primordiale"
of Italian MIUR.


\end{document}